\documentclass{article}
\usepackage{spconf,amsmath,graphicx,hyperref}
\usepackage{amssymb}
\usepackage{enumitem}
\usepackage{booktabs}
\usepackage{xcolor}
\hypersetup{hidelinks}

\def\L{{\cal L}}

\title{Real-Time Hybrid Performance-MIDI-to-Score Transcription \\ with 
Offline-to-Online Knowledge Distillation}
\name{Lynn Ye \qquad Roger Dannenberg \qquad Chris Donahue}
  \address{Carnegie Mellon University
  }

\begin{document}
%
\maketitle
\begin{abstract}
Turning a piano performance into conventional music notation requires recovering score structure from expressive performance timing. Here, we study performance-MIDI-to-score (PM2S) transcription, in which note pitches and timings captured by a digital instrument are converted into a musical score. Existing PM2S systems rely on the complete performance, precluding real-time applications such as live score visualization and interactive composition. We formulate strictly causal, event-synchronous PM2S and introduce an online model that updates the score after every note with zero lookahead and 1.75\,ms average latency on consumer hardware. We further propose note-aligned knowledge distillation from an offline teacher, reducing the online model's average transcription error from 13.49\% to 12.65\% under standard PM2S evaluation, approaching the teacher's 11.39\%. Their shared score representation enables a hybrid system that combines immediate online predictions with periodic offline refinement, achieving the best of both worlds\footnote{Examples:  \url{https://lynnzye.github.io/OPM2S-demo}}.

\end{abstract}
\section{Introduction}
\label{sec:intro}
Automatic music transcription (AMT) converts musical performances into symbolic representations for editing, analysis, and visualization \cite{amt}. This broad family of tasks includes audio-to-MIDI \cite{tkun}, audio-to-score \cite{zeng24}, and performance-MIDI-to-score (PM2S) \cite{liu22, beyer24}. PM2S isolates the notation problem: given pitches and performance timings, it infers their quantized positions and organization in a musical score. This is challenging because expressive timing and tempo variations obscure the metrical grid, while the system must infer numerous interdependent score attributes, including bar lines, meter, beats, staff, and voice.

\begin{figure}
  \centering
  \includegraphics[width=0.95\linewidth]{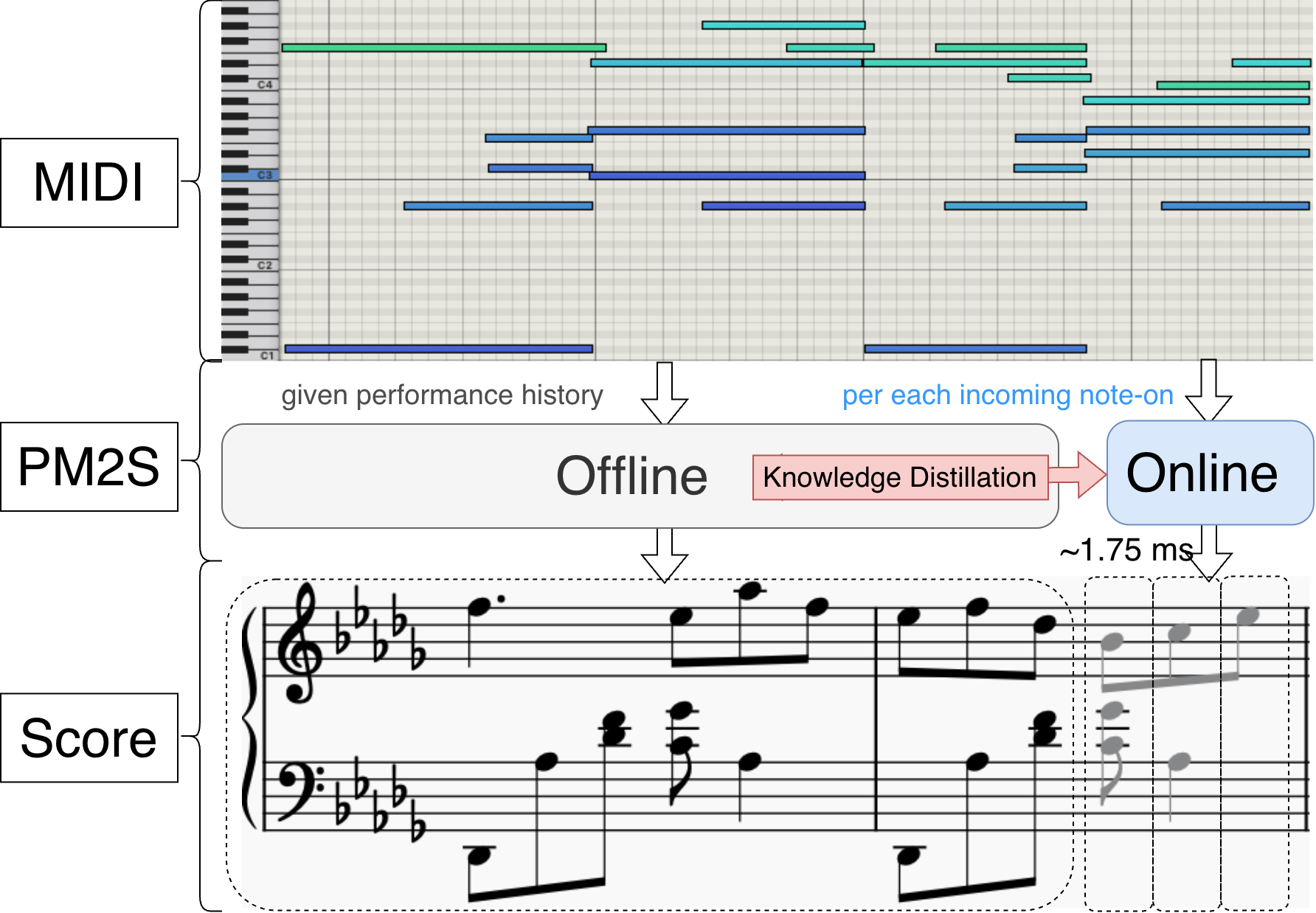}
  \caption{Our real-time PM2S system features hybrid online and offline models. The online model updates the score after each note, while the offline model revises history as context becomes available. Distillation 
  improves the online model.
  }
  \label{fig:fig1}
\end{figure}

Existing PM2S systems use complete performances or large chunks to resolve ambiguities in timing and score structure. However, applications such as live score visualization and interactive composition require predictions with low latency and without future context, as in analogous streaming scenarios in speech recognition \cite{Misra2021, kojima21_interspeech}.

Our framework uses a shared note-aligned representation across two complementary models: a decoder-only model provides immediate predictions, while an encoder--decoder model uses available context to revise both the score and the online model's decoding history. The offline model additionally transfers its future-informed representations to the online model through matched-history knowledge distillation.  

Our key contributions are:
\begin{itemize}[noitemsep, topsep=0pt]
    \item We are the first to formulate strictly causal, event-synchronous PM2S, and introduce a decoder-only model that predicts one score token per note with zero lookahead and 1.75\,ms average inference time, while incurring only 18.4\% higher average MUSTER error than our offline model. 
    \item We introduce note-aligned knowledge distillation from an encoder--decoder teacher, further reducing the error gap to 11.1\% without increasing inference time. 
    \item We propose a hybrid system where offline predictions periodically refine the score and causal decoding history without representation translation. Conditioning on one offline-refined prefix preserves average MUSTER error while reducing subsequent onset and staff errors by 5.0\% and 16.2\%,  respectively.
\end{itemize}

\section{Related Work}

\subsection{Score-Level Music Transcription}
Score-level AMT follows two main paradigms: cascaded audio-to-MIDI-to-score (A2M2S) and end-to-end audio-to-score (A2S). A2M2S systems first estimate the pitches and note timings from audio \cite{tkun, Nakamura18}, then convert the resulting performance MIDI into a notated score. The second stage, known as performance-MIDI-to-score (PM2S), has been approached using graphical models \cite{Nakamura18}, recurrent networks \cite{liu22}, and encoder--decoder Transformers \cite{beyer24}. Concurrent work introduces DuET, an offline PM2S model with joint frame-level metrical modeling \cite{zeng2026duet}, released after our experiments were completed and thus not included in our comparison. These methods operate on complete performances or large chunks of notes and are therefore designed for offline transcription. 

In contrast, we study whether PM2S can be performed causally as each MIDI event arrives. Our formulation maintains a one-to-one correspondence between performed notes and predicted score tokens, enabling event-synchronous inference, aligned knowledge distillation from an offline model, and a hybrid inference framework. 

Additionally, others have explored A2S systems in an offline setting, generating scores directly from audio \cite{zeng24, liu21}. While this approach avoids an intermediate representation, the model must jointly recognize notes and infer score structure, without necessarily maintaining correspondence between acoustic events and score tokens.

\subsection{Streaming and Knowledge Distillation}
Knowledge distillation (KD) from offline teachers has improved streaming recognition through data generation \cite{Misra2021}, output-distribution matching \cite{hinton2015distillingknowledgeneuralnetwork}, and hidden-state alignment \cite{kojima21_interspeech}. Prior latent-distillation methods typically align corresponding layers of structurally similar teacher and student models \cite{Shim2023, tang23}. In music information retrieval, HKDSME \cite{kdhetero} transfers intermediate features between heterogeneous networks for lightweight offline singing-melody extraction. More closely related to our offline-to-online setting, ReaLChords distills a future-aware encoder--decoder model to a causal decoder-only chord model, but uses output-level, on-policy KL regularization \cite{realchord}. We instead match note-aligned task-facing hidden states between offline encoder--decoder and online decoder-only PM2S models.

\section{Methodology}
\label{sec:method}

Online PM2S updates the score after each note onset using only observations available at that time. This setting presents three challenges: note durations are unavailable at onset, measure-level attributes require evidence across multiple events, and autoregressive inference compounds errors. We address these challenges using compact note-aligned score representations, masked-history training, and knowledge distillation from a full-context teacher.

\subsection{Event-Synchronous PM2S and Score Representation}
Let a MIDI performance be an onset-ordered sequence
\begin{equation*}
X=(x_1,\ldots,x_N),\quad
x_i=(p_i,\Delta_i),\quad
\Delta_i=\tau_i-\tau_{i-1},
\end{equation*}
where $p_i$ is the MIDI pitch and
$\Delta_i$ is the inter-onset interval between note onset times $\tau_i$ and $\tau_{i-1}$, where $\tau_i \geq \tau_{i-1}$. Simultaneous notes have $\Delta_i=0$ with ordering ties broken by pitch. We retain the full MIDI pitch range and logarithmically quantize $\Delta_i$ over the interval from 0 to 10\,s into 257 bins.
\begin{equation*}
\Delta'_i = \min\left(\bigg\lfloor \frac{\log(1+\min(\Delta_i, 10)/0.5)}{\log(1+10/0.5)} \cdot 256 \bigg\rfloor,256\right)
\end{equation*}
We define an \emph{event-synchronous} PM2S system as producing one causal score update at each note onset without future context. Attributes unavailable at onset may take provisional values and be revised later.

Because complete notation includes attributes unavailable at note onset, we predict a minimal note-level state sufficient for constructing a reasonable score: metrical positions and staff assignment. We represent metrical position using \emph{tactus}, the predominant metrical pulse, and \emph{tatum}, its subdivision. Let $a_i\in \{0,\ldots,4\}$ be the tactus position and $u_i\in\{0,\ldots,23\}$ be its subdivision. Their combination, $t_i=24a_i+u_i$, defines 120 \emph{tactum} tokens plus one additional token $t_i=120$ representing irregular events such as grace notes. Compound meters use their canonical tactus counts (e.g., two for 6/8), and unsupported rare meters such as 7/8 are excluded only from training. Together with a staff label $s_i\in\{0,1\}$ representing the left- and right-hand staves, respectively, the prediction target is $y_i=(t_i,s_i)$. MIDI pitch is copied directly, while the remaining score structure is derived during postprocessing.

The offline $P_{\text{off}}$ and online $P_{\text{on}}$ factorizations differ only in their access to performance context:
\begin{align}
P_{\mathrm{on}}(Y\mid X)
&=\prod_{i=1}^{N}P_{\mathrm{on}}(y_i\mid x_{\leq i},y_{<i})\label{eq:online}\\
P_{\mathrm{off}}(Y\mid X)
&=\prod_{i=1}^{N}P_{\mathrm{off}}(y_i\mid X,y_{<i})
\label{eq:offline}
\end{align}

\subsection{Online and Offline PM2S Models}
\label{sec}
We propose to model Eq.~\ref{eq:online} using a decoder-only Transformer with rotary positional embeddings (RoPE) \cite{su2021roformer}. At position $i$, it receives the sum of embeddings $\psi_i$ for the current pitch $p_i$, the current inter-onset interval $\Delta_i$, the previously generated tactum $t_{i-1}$, and the previous staff label $s_{i-1}$:
\begin{equation*}
e_i =
\psi_p(p_i)+\psi_{\Delta}(\Delta_i)
+\psi_q(t_{i-1})+\psi_s(s_{i-1})
\label{eq:emb}
\end{equation*}

\begin{figure}
  \centering
  \includegraphics[width=0.85\linewidth]{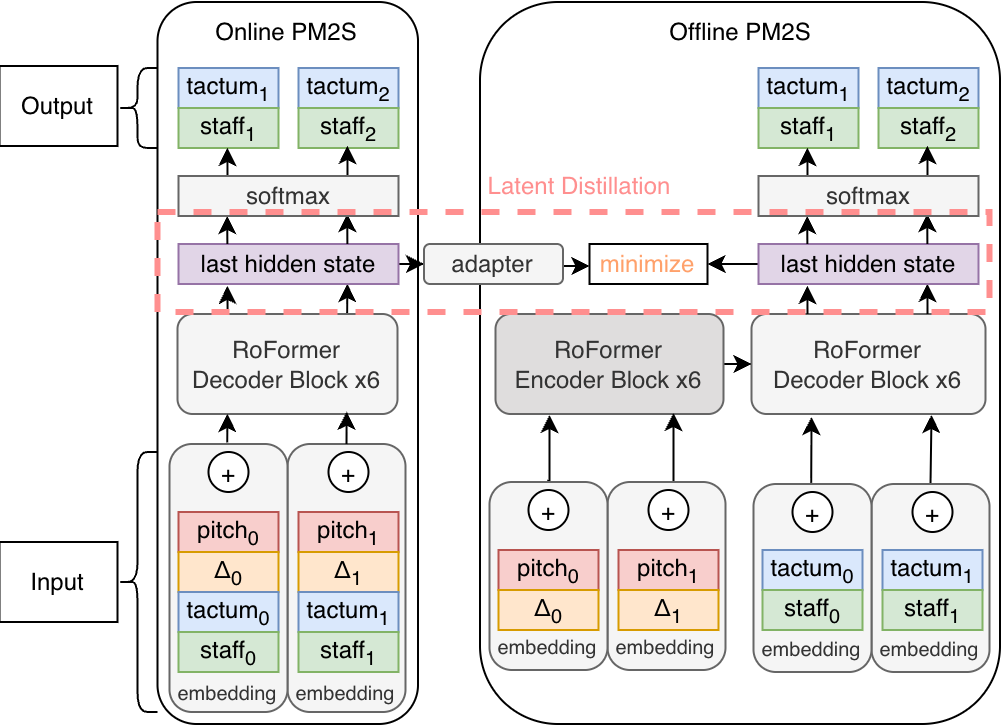}
  \caption{A visualization of our proposed model architecture. We use a hidden dimension of 256, a feed-forward dimension of 768, and 8 attention heads.}
  \label{fig:fig2}
\end{figure}

Causal self-attention ensures the prediction depends only on $x_{\leq i}$ and $y_{<i}$. Two linear heads predict tactum and staff.

We parameterize Eq.~\ref{eq:offline} with an encoder--decoder Transformer with RoPE, predicting the same targets. Its bidirectional encoder processes the complete sequence of pitch and inter-onset-time token embeddings $E_{\text{off}}=(e'_1,\ldots,e'_N)$:
\begin{equation*} 
    e'_i=\psi_p(p_i)+\psi_\Delta(\Delta_i)
\end{equation*}
A causal decoder receives the preceding tactum and staff labels and cross-attends to every encoded performance event. Details of the model architecture are included in Fig.~\ref{fig:fig2}.

The shared note-aligned score representation establishes a direct correspondence, enabling hybrid inference (\ref{sec:hybrid}) and a note-aligned knowledge distillation method (\ref{sec:latentkd}).

\subsection{Hybrid System}
\label{sec:hybrid}
We propose a hybrid system that combines immediate causal transcription with periodic offline refinement. After every $S$ new notes, the offline model processes the latest window of $L$ notes, using the first $L-S$ overlapping predictions as context, and replaces the predictions for the newest $S$ notes. During this pass, the online model continues without interruption. Once the refined history is available, a separate causal instance replays any notes that arrived in the meantime and, upon catching up, replaces the live decoding state. Thus, each note receives an immediate zero-lookahead prediction, while earlier predictions and their causal history are progressively refined. A final offline pass revises the remaining suffix, making the completed transcription equivalent to offline inference. The window length $L$ and stride $S$ control the trade-off among available context, refinement frequency, and computation. In practice, we follow \cite{beyer24} and set $L=512$ during training and inference, and set $S=256$ (50\% overlap) to strike a balance between quality and latency. 

\subsection{Note-Aligned Knowledge Distillation}\label{sec:latentkd}
Prior offline-to-streaming distillation methods transfer future-context information by aligning hidden representations of teacher and student models \cite{kojima21_interspeech,Shim2023,tang23}. Building on this principle, we introduce note-aligned representation distillation between our heterogeneous PM2S architectures. Since both models predict the same target for every performed note, their task-facing states have an exact temporal correspondence and can be aligned directly at every note. 

Let $h_i^{\mathrm{on}}=f_\theta(x_{\le i}, y_{<i})\in\mathbb{R}^{D}$ be the online model's final hidden state at timestep $i$, and $h_i^{\mathrm{off}}~=~g_\phi(X, y_{<i})\in\mathbb{R}^{D}$ denote the corresponding frozen offline model's final decoder state. We additionally transform the student state using a learned linear adapter $A\in\mathbb{R}^{D\times D}$, initialized to the identity matrix. Following representation-level distillation \cite{kojima21_interspeech}, the token-level knowledge distillation objective is:
\begin{equation}
\mathcal{L}_{\text{lat}}
=
\frac{1}{|\mathcal{V}|D}
\sum_{i\in\mathcal{V}}\sum_{d=1}^{D}
\rho\left(
[Ah_i^{\mathrm{on}}]_d -
\operatorname{sg}\left([h_i^{\mathrm{off}}]_d\right)
\right)
\label{eq:distill}
\end{equation}
where $\mathcal{V}$ is the set of non-padding note positions, $\operatorname{sg}$ denotes stop-gradient, and $\rho$ is the Smooth-$L_1$ penalty.

The supervised objective averages the tactum and staff cross-entropies. The student objective adds $\lambda \L_{\text{lat}}$ to the mean tactum and staff cross-entropies. We also compare our method with output-distribution KL matching and layerwise knowledge distillation in our ablations.
\par\smallskip\noindent
\textbf{Matched-History Masking} \quad Teacher-forced training exposes the model only to ground-truth histories, causing error propagation during inference. Following \cite{beyer24}, we mask each preceding score label with 75\% probability. During distillation, the same sampled mask is applied to both models, preventing the teacher from receiving a more informative score history than the student. 

\subsection{Post Processing}\label{sec:pp}
Predicted tactums define a metrical grid from which we derive the remaining attributes needed to render a score. Note values are obtained by mapping MIDI note-offs to their nearest quantized grid positions. Measure boundaries and time signatures are inferred from tactus positions and phase resets, while key signatures are estimated from the accumulated performance using the Krumhansl--Schmuckler algorithm \cite{krumhansl1990cognitive}.

\section{Experiments}
\begin{table*}[t]
\centering
\resizebox{\textwidth}{!}{
\begin{tabular}{llrrrrrrrrrrrrr}
\toprule
& &
\multicolumn{6}{c}{MUSTER \cite{Nakamura18, Hiramatsu2021-af}} &
\multicolumn{3}{c}{ScoreSimilarity \cite{cogliati2017metric}} &
\multicolumn{1}{c}{RTF} &
\multicolumn{1}{c}{Latency} &
\multicolumn{2}{c}{Structural delay} \\
\cmidrule(lr){3-8}
\cmidrule(lr){9-11}
\cmidrule(lr){12-15}

System & Mode &
Pitch$\downarrow$ & Miss$\downarrow$ & Extra$\downarrow$ &
Onset$\downarrow$ & Offset$\downarrow$ & Avg.$\downarrow$ &
Staff$\downarrow$ & Key$\downarrow$ & Meter$\downarrow$ &
Avg.$\uparrow$ &
ms/note$\downarrow$ &
Mean (s)$\downarrow$ &
P95 (s)$\downarrow$ \\
\midrule

Liu et al. \cite{liu22}
& Offline
& $\mathbf{2.02}$
& $\underline{6.81}$
& ${9.01}$
& $68.28$
& $54.11$
& $28.04$
& $6.86$
& $24.63$
& $53.03$
& $\mathbf{128.8}$
& $\mathbf{0.73}$
& $186.56$
& $422.76$ \\

Beyer et al. \cite{beyer24}
& Offline
& $3.11$
& $7.56$
& $\mathbf{6.44}$
& $\underline{15.55}$
& $\mathbf{23.84}$
& $\mathbf{11.30}$
& $\underline{6.62}$
& $45.46$
& $52.59$
& $14.8$
& $7.55$
& ${21.94}$
& ${61.62}$ \\

\textbf{Ours Online}
& Online
& $2.21$
& $8.25$
& $9.59$
& $16.74$
& $30.66$
& $13.49$
& $7.26$
& $\mathbf{14.90}$
& $\mathbf{28.18}$
& $\underline{57.4}$
& $\underline{1.75}$
& $\mathbf{0.00}$
& $\mathbf{0.00}$ \\

\textbf{Ours Online + KD}
& Online
& $\underline{2.17}$
& $7.13$
& $8.96$
& $16.12$
& $28.84$
& $12.65$
& $6.82$
& $\underline{15.12}$
& $\mathbf{28.92}$
& $\underline{57.4}$
& $\underline{1.75}$
& $\mathbf{0.00}$
& $\mathbf{0.00}$ \\

\textbf{Ours Offline}
& Offline
& $\underline{2.17}$
& $\textbf{6.26}$
& $\underline{8.49}$
& $\mathbf{13.25}$
& $\underline{26.77}$
& $\underline{11.39}$
& $\mathbf{3.27}$
& ${20.43}$
& $\underline{38.73}$
& $7.91$
& $12.71$
& $\underline{14.63}$
& $\underline{45.61}$ \\

\textbf{Ours Hybrid}
& Hybrid
& $\underline{2.17}$
& $\textbf{6.26}$
& $\underline{8.49}$
& $\mathbf{13.25}$
& $\underline{26.77}$
& $\underline{11.39}$
& $\mathbf{3.27}$
& ${20.43}$
& $\underline{38.73}$
& $\underline{57.4}$
& $\underline{1.75}$
& $\mathbf{0.00}$
& $\mathbf{0.00}$ \\

\bottomrule
\end{tabular}
}
\caption{PM2S results on the test set proposed by Beyer et al.~\cite{beyer24}.
Results for \cite{liu22} and \cite{beyer24} are quoted from \cite{beyer24} except for metrics not measured. Runtime measurements were obtained on a MacBook Pro M1 with batch size 1.
RTF: performance duration divided by inference time. Structural delay: time to obtain enough context for inference. Our model results are averaged across four random seeds. Transformer-based models use KV-cached inference with no warm-up. The hybrid system's RTF and latency describes only the immediate online path. Its error rates are measured after the final offline refinement.
}
\label{tab:comparison}
\end{table*}
\label{sec:experiments}
\subsection{Setup}
\textbf{Dataset}$\quad$ We train on PianoCoRe \cite{borovik2026pianocore}, which provides note-level alignments between score and performance MIDI. We heuristically align MusicXML notes to the score MIDI to extract notation attributes, discard poorly aligned pieces, and remove certain test pieces from training and validation, following \cite{beyer24}. No score appears in two splits. The resulting train, validation, and test splits contain 1,206~/~77~/~80 scores and 83,850~/~8,272~/~13,064 performances, respectively. 
\par\smallskip\noindent
\textbf{Training}$\quad$ All models are trained for 15,100 steps on randomly sampled segments of up to 512 notes using AdamW \cite{loshchilov2019decoupled}, a learning rate of $3{\times}10^{-4}$, batch size 8, linear warmup, and cosine decay. We augment with pitch shifts in $[-5,+6]$ semitones, tempo scaling by $[0.8, 1.2]$, and onset jitter. The distillation weight $\lambda=2$ is tuned on the validation set. 

\subsection{Evaluation}
We evaluate the final accumulated score with standard PM2S metrics: MUSTER \cite{Nakamura18, Hiramatsu2021-af} and ScoreSimilarity \cite{cogliati2017metric}. Both align the generated and reference scores before computing attribute-level error rates. All score updates use only the available performance context. For online-only evaluation, each tactum and staff token is predicted once at note onset and never revised. Postprocessing updates only locally unresolved attributes such as note duration and meter. 
Table~\ref{tab:comparison} shows that our offline model is competitive with the previous state of the art \cite{beyer24} despite predicting a more restricted score representation. Knowledge distillation reduces the online model's average MUSTER error from 13.49\% to 12.65\% while improving most individual metrics. On a MacBook Pro M1, the online model predicts each score token in 1.75\,ms (P95: 1.84\,ms) with zero structural delay for tactum and staff tokens, supporting event-synchronous PM2S. 
\par\bigskip\noindent
\textbf{Ablation Study}\quad Table~\ref{tab:ablation} compares variants of the online model. Shared
Masking (SM) applies the same masked decoder history to the teacher and student. Our proposed knowledge distillation improves average MUSTER error and reduces variance across seeds relative to training without distillation, with stronger alignment weights generally performing better. Removing shared masking (No SM) from the offline model during KD substantially degrades performance, while KL and layerwise distillation underperform final-state alignment. Decoder-history masking is crucial, reducing average MUSTER error by approximately seven percentage points. 
\par\bigskip\noindent
\textbf{Hybrid System} \quad Table~\ref{tab:hybrid} 
examines whether conditioning the online model on the offline-refined history improves its performance relative to conditioning on its own predictions. 
Incorporating a single 256-note offline prefix preserves average MUSTER error while reducing onset and staff errors by 5.0\% and 16.2\%, respectively. These results show that offline revisions are compatible with causal continuation and may improve subsequent prediction quality. 

\label{sec:hybrideval}
\begin{table}[t]
\centering

\resizebox{\columnwidth}{!}{
\begin{tabular}{lrrrrr}
\toprule
Setting &
Miss$\downarrow$ &
Extra$\downarrow$ &
Onset$\downarrow$ &
Avg.$\downarrow$ &
Staff$\downarrow$ \\
\midrule

SM $\lambda=2$
& $\mathbf{7.1{\pm}0.7}$
& $\mathbf{9.0{\pm}0.3}$
& $\mathbf{16.1{\pm}0.3}$
& $\mathbf{12.7{\pm}0.3}$
& $6.8{\pm}2.1$ \\
\midrule

SM $\lambda=1$
& $8.4{\pm}1.8$
& $9.7{\pm}1.0$
& $\underline{16.5{\pm}0.7}$
& $\underline{13.4{\pm}1.0}$
& $6.9{\pm}1.5$ \\

SM $\lambda=0.5$
& $10.5{\pm}5.8$
& $10.5{\pm}2.5$
& $17.0{\pm}2.1$
& $14.3{\pm}2.9$
& $6.1{\pm}2.0$ \\

No KD
& $\underline{8.3{\pm}1.0}$
& $\underline{9.6{\pm}0.4}$
& $16.7{\pm}0.6$
& $13.5{\pm}0.5$
& $7.3{\pm}3.7$ \\



\addlinespace

\addlinespace
KL loss
& $10.2{\pm}3.6$
& $10.5{\pm}1.5$
& $16.6{\pm}1.2$
& $14.0{\pm}1.7$
& $5.1{\pm}0.5$ \\

Layerwise KD
& $11.3{\pm}3.7$
& $10.6{\pm}1.5$
& $16.9{\pm}1.8$
& $14.4{\pm}1.8$
& $5.7{\pm}0.6$ \\

\addlinespace
No SM
& $20.7{\pm}4.0$
& $15.3{\pm}1.5$
& $20.1{\pm}0.5$
& $19.0{\pm}1.4$
& $\mathbf{4.4{\pm}0.4}$ \\

No 75\% mask
& $21.7{\pm}3.7$
& $16.1{\pm}1.3$
& $21.1{\pm}1.3$
& $19.8{\pm}1.7$
& $\underline{4.8{\pm}0.9}$ \\

\bottomrule
\end{tabular}
}

\caption{Values are mean $\pm$ standard deviation across four seeds. All
metrics are error rates in percent; lower is better.}
\label{tab:ablation}
\end{table}
\begin{table}[t]
\centering

\resizebox{\columnwidth}{!}{
\begin{tabular}{lrrrrr}
\toprule
History &
Miss$\downarrow$ &
Extra$\downarrow$ &
Onset$\downarrow$ &
Avg.$\downarrow$ &
Staff$\downarrow$ \\
\midrule

Offline Prefix
& $\mathbf{7.0{\pm}0.5}$
& $\mathbf{8.9{\pm}0.3}$
& $\mathbf{15.3{\pm}0.5}$
& $\mathbf{12.6{\pm}0.4}$
& $\mathbf{5.7{\pm}0.6}$ \\


Online Estimated
& ${7.1{\pm}0.7}$
& $9.0{\pm}0.3$
& $16.1{\pm}0.3$
& $12.7{\pm}0.3$
& $6.8{\pm}2.1$ \\

\bottomrule
\end{tabular}
}

\caption{Offline Prefix conditions the online model on offline predictions for the first 256 notes. Online Estimated uses only online inference results. 
}
\label{tab:hybrid}
\end{table}

\section{Conclusion}
\label{sec:conclusion}
We introduced strictly causal, event-synchronous PM2S with 1.75\,ms average per-note inference. Note-aligned distillation reduced average MUSTER error from 13.49\% to 12.65\% for the online model, while offline-refined histories enabled a hybrid framework with the best of both worlds: immediate transcription and offline-quality scores after history refinement. 
Code will be released upon publication.

\vfill\pagebreak

\ninept
\bibliographystyle{IEEEbib}
\bibliography{strings,refs}

\end{document}